\documentclass[pra, aps, twocolumn, floatfix, showpacs]{revtex4-1}
\usepackage{graphicx, amsmath, amssymb, times}

\begin{document}
\title{Two-band superfluidity and intrinsic Josephson effect \\ 
in alkaline-earth Fermi gases across an orbital Feshbach resonance}
\author{M. Iskin}
\affiliation{
Department of Physics, Ko\c c University, Rumelifeneri Yolu, 34450 Sar{\i}yer, Istanbul, Turkey.
}
\date{\today}
\begin{abstract}

We first show that the many-body Hamiltonian governing the physical
properties of an alkaline-earth $^{173}$Yb Fermi gas across the
recently-realized orbital Feshbach resonance is exactly analogous to that 
of two-band $s$-wave superconductors with contact interactions: 
\textit{i.e.}, even though the free-particle bands have a tunable energy 
offset in between and are coupled by a Josephson-type attractive inter-band 
pair scattering, the intra-band interactions have exactly the same strength. 
We then introduce two intra-band order parameters within the BCS 
mean-field approximation, and investigate the competition between their 
in-phase and out-of-phase (\textit{i.e.}, the so-called $\pi$-phase) solutions 
in the entire BCS-BEC evolution at zero temperature.

\end{abstract}
\pacs{03.75.Ss, 03.75.-b, 03.75.Hh, 03.75.Mn}
\maketitle

\textit{Introduction.}
Over the past decade or so, the cold-atom systems have emerged as 
versatile quantum simulators of few- and many-body physics 
theories~\cite{chin10, bloch08, giorgini08, dalibard11, galitski13}.
Thanks partly to their high degree of tuning capacities, they proved to
be ideal test beds for exploring not only few-body phenomena including 
the simplest two-body and exotic Efimov bound states~\cite{chin10, blume12} 
but also macroscopic phases of matter ranging from BEC and 
superfluidity of atomic Bose gases, BCS superfluidity and BCS-BEC 
crossover of atomic Fermi gases, superfluid-Mott insulator transition 
and quantum magnetism of lattice gases, and topological insulators and 
superfluids~\cite{bloch08, giorgini08, dalibard11, galitski13}. In addition, 
atomic systems also bridge the gap between the macroscopic properties 
of many-body systems to microscopic physics of their constituent 
particles under the same setting, providing hindsights into the realm 
of mesoscopic systems as well~\cite{wenz13, zurn13}.

One of the most crucial ingredients behind this success is the ability to
control the strength and symmetry of the inter-particle interactions with 
atomic precision~\cite{chin10}. In particular, since alkali atoms 
(with a single valence electron) have nonzero electronic 
angular momentum in their ground states, they are highly sensitive to 
externally-applied magnetic fields, and the Zeeman shifts of their different 
electronic-spin states provide a knob to control the inter-hyperfine-state 
interactions. More specifically, alkali atoms allow for tunable couplings 
between the energies of a two-body closed-channel bound state and 
of two interacting open-channel atoms via what is known as the 
magnetic Feshbach resonance~\cite{chin10}.
On the other hand, while the zero electronic angular momentum of 
$^{173}$Yb-like alkaline-earth atoms (with two valence electrons) 
make them highly insensitive to external magnetic fields, it is still 
possible to control the inter-orbital-state interactions with only tens 
of Gauss via what is known as the orbital Feshbach 
resonance~\cite{zhang15, pagano15, hofer15, cornish15, xu16}. 
Here, it is the Zeeman shifts of different nuclear-spin states of atoms 
that is used to tune the coupling between closed and open channels. 
Given that the experimentalists are now pursuing towards the superfluid 
regime~\cite{pagano15, hofer15, cornish15}, the newly-realized orbital 
Feshbach resonance in a $^{173}$Yb Fermi gas promise a new wave of inter-disciplinary 
interest for studying many-body phenomena in an uncharted atomic-physics 
territory due to its direct connection to a wide-class of so-called 
two-band superconductors, which includes many of the recently discovered 
iron-based superconductors~\cite{tsuda03, souma03, hunte08, ruby15}.

For instance, it turns out that the many-body Hamiltonian governing 
the physical properties of alkaline-earth Fermi gases across an 
orbital Feshbach resonance~\cite{zhang15, xu16} is exactly 
analogous to that of two-band $s$-wave superconductors 
with contact interactions in one of its most simplest 
forms~\cite{suhl59, iskin05, iskin06, iskin07}. That is, even though the 
free-particle bands are shifted by a tunable energy offset in between and 
are coupled by an attractive inter-band pair scattering, the intra-band 
interactions have exactly the same strength. In a broader context, 
since the two-band Hamiltonian also resembles to that of Josephson 
junctions between two condensates, the physics across an orbital 
Feshbach resonance can also be thought of as an intrinsic Josephson 
effect~\cite{leggett66, kleiner94, moll14}. The key role played by the inter-band 
Josephson coupling between two intra-paired bands is played in 
atomic systems by the intra-orbital scattering lengths between the 
open- and closed-channel atoms. Thus, motivated by the recent 
theoretical and experimental 
proposals~\cite{zhang15, pagano15, hofer15, cornish15, xu16}, 
here we introduce two intra-band order parameters within the BCS 
mean-field approximation, and investigate the competition between 
their in-phase and out-of-phase solutions in the entire BCS-BEC 
evolution at zero temperature.

\begin{figure}[htb]
\centerline{\scalebox{0.3}{\includegraphics{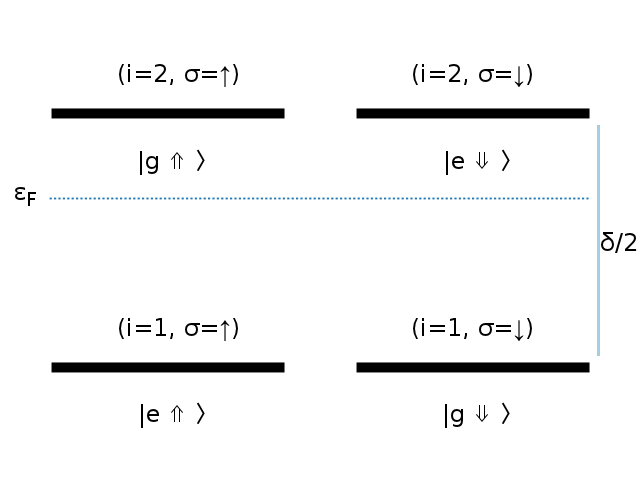}}}
\caption{\label{fig:bands} (Color online)
Coupling between $^{173}$Yb atoms in two orbital states ($g$ and $e$)
with two nuclear-spin projections ($\Uparrow$ and $\Downarrow$) can 
effectively be described as a pseudo-spin-1/2 ($\uparrow$ and $\downarrow$) 
Fermi gas with two non-degenerate bands ($1$ and $2$). Here, the 
band-offset $\delta/2$ is a tunable parameter, and the Fermi energy 
$\varepsilon_F$ is the energy scale used in numerics.
}
\end{figure}

\textit{Effective Two-band Model.}
To describe alkaline-earth Fermi gases across an orbital Feshbach 
resonance, we start with the momentum-space Hamiltonian~\cite{footHam}
\begin{align}
H = \sum_{i \sigma \mathbf{k}} \xi_{i \mathbf{k}} c_{i \sigma \mathbf{k}}^\dagger  c_{i \sigma \mathbf{k}}
- \sum_{i j \mathbf{q}} V_{ij} b_{i \mathbf{q}}^\dagger b_{j \mathbf{q}},
\label{eqn:ham}
\end{align}
where the band index $i \equiv \{1, 2\}$ and pseudo-spin index $\sigma \equiv \{\uparrow, \downarrow\}$ 
correspond, respectively, to four-way superpositions of atoms in two different 
orbital states $\{g, e\}$ with two possible nuclear-spin projections 
$\{\Uparrow, \Downarrow\}$. As summarized in Fig.~\ref{fig:bands}, the $i = 1$ 
and $2$ bands refer, respectively, to the atoms in the open and closed channels. 
The operator $c_{i \sigma \mathbf{k}}^\dagger$ creates a single $\sigma$-particle 
in band $i$ with momentum $\mathbf{k}$ and dispersion
$
\xi_{i \mathbf{k}} = \varepsilon_{\mathbf{k}} - \mu_i,
$
where $\varepsilon_{\mathbf{k}} = k^2/(2m)$ is the energy of a free particle 
(in units of $\hbar = 1$), and the energy shift (\textit{i.e.}, the detuning 
parameter $\delta \ge 0$) is incorporated into the effective chemical 
potentials as $\mu_1 = \mu$ and $\mu_2 = \mu - \delta/2$.

Since the density-density interactions are typically short-ranged in atomic systems, 
the zero-ranged interactions accounted by the second term in Eq.~\ref{eqn:ham} 
may be sufficient for our purposes, where the operator
$
b_{i \mathbf{q}}^\dagger = \sum_{\mathbf{k}} 
c_{i \uparrow, \mathbf{k}+\mathbf{q}/2}^\dagger  
c_{i \downarrow, -\mathbf{k}+\mathbf{q}/2}^\dagger
$
creates pairs of $\uparrow$ and $\downarrow$ particles in band $i$ with 
center-of-mass momentum $\mathbf{q}$. While the intra-band interactions turned
out to be attractive with equal amplitudes
$
V_{11} = V_{22} = (g_- + g_+)/2 \ge 0,
$
the inter-band interaction
$
V_{12} = V_{21} = (g_- - g_+)/2
$
may be attractive (positive) or repulsive (negative) depending on the particular 
species of atom at hand. Here, $g_\pm \ge 0$ are the amplitudes for the bare 
atom-atom attractions in the singlet/triplet (anti-symmetric/symmetric) 
superpositions of the orbital degrees of freedom, and they are related to the 
corresponding scattering lengths $a_{s\pm}$ via the usual renormalization relations
$
1/g_\pm = -m \Omega/(4\pi a_{s\pm}) + \sum_{\mathbf{k}} m/k^2.
$
Here, $\Omega$ is the volume, and even though both $g_\pm$ and $V_{ij}$
depend explicitly on the particular value of cutoff ($k_0$) used in $\mathbf{k}$-space
integrals, our many-body results given below are independent of $k_0$ as 
long as $k_0 \to \infty$ is chosen sufficiently high. Therefore, since
$V_{12} \gtrless 0$ for $1/a_{s-} \gtrless 1/a_{s+}$, we note that $V_{12}$ is 
positive for a $^{173}$Yb Fermi gas where $a_{s+} \approx 1900a_0$ and 
$a_{s-} \approx 200a_0$ with $a_0$ the Bohr radius~\cite{pagano15, hofer15}.

\textit{Two-Body Bound States in Vacuum.}
It has recently been shown that the two-body $s$-wave scattering length 
between two open-channel alkaline-earth atoms is given by~\cite{zhang15},
$
a_s = a_{s2} + \sqrt{m \delta} a_{s1}^2 
       / (1 - a_{s2} \sqrt{m \delta}),
$
where $a_{s2(1)} = (a_{s-} \pm a_{s+})/2$. This shows that $a_s$ changes 
from $a_{s2}$ to $a_{s+} a_{s-}/a_{s2}$ as $\delta$ increases from $0$ to 
$\infty$, with an intermediate divergence at $\delta_{res} = 1/(m a_{s2}^2)$ 
when the conditions $a_{s2} > 0$ and $a_{s1} \ne 0$ are both satisfied, 
and an intermediate zero crossing at 
$\delta_0 = 1/[m(a_{s2}-a_{s1}^2/a_{s2})^2]$ when the condition 
$a_{s2} > |a_{s1}|$ is satisfied. Note that since $a_s = a_{s1}^2/(a_{s2}0^\pm)$
as $\delta \to \delta_{res}^\mp$, there are two possible scenarios when 
$a_{s2} > 0$ and $a_{s1} \ne 0$.
In the first-case scenario, if both $a_{s+}$ and $a_{s-}$ are positive then 
increasing $\delta$ first increases $a_s$ from $a_{s2}$ to $+\infty$ and then 
from $-\infty$ to $a_{s+} a_{s-}/a_{s2} > 0$ with an intermediate zero crossing 
at $\delta_0  > \delta_{res}$. This is the case for a $^{173}$Yb Fermi 
gas for which $a_{s2} > |a_{s1}| > 0$.
In the second-case scenario, if only one of $a_{s\pm}$ is negative in such 
a way that $0 < a_{s2} < |a_{s1}|$ then increasing $\delta$ first increases 
$a_s$ from $a_{s2}$ to $+\infty$ and then from $-\infty$ to $a_{s+} a_{s-}/a_{s2} < 0$ 
without an intermediate zero crossing.

It is well-known that a two-body bound state in vacuum is characterized 
by $a_s > 0$ with binding energy $\varepsilon_b = 1/(m a_s^2)$. This suggests 
that increasing $\delta$ from $0$ to $\delta_{res}^-$ gradually weakens 
the binding of atoms as $a_s$ increases from $a_{s2}$ to $+\infty$,
beyond which the divergence of $a_s \to -\infty$ as $\delta \to \delta_{res}^+$ 
signifies the complete unbinding of atoms, \textit{i.e.}, the disappearance 
of the two-body bound state. In addition, since $a_s \to a_{s2}$ not only in the 
$\delta \to 0$ but also in the $a_{s1} \to 0$ limit, the conditions $\delta > 0$, 
$a_{s1} \ne 0$ and $a_{s2} > 0$ are all essential requirements for 
realizing an orbital Feshbach resonance. For instance, it turns out that 
a $^{173}$Yb Fermi gas with $a_{s+} \gg a_{s-} \gg a_0$ requires a so low 
$\delta_{res}$ threshold that it has recently allowed for the very first creation 
of such a resonance shortly after its theoretical 
prediction~\cite{zhang15, pagano15, hofer15, cornish15}. 
Even though two-body bound states are not allowed in the open channel
when $a_s \le 0$ (or equivalently $\delta \ge \delta_{res}$), many-body 
bound states may still prevail in the ground state as Cooper pairs, 
which is discussed next.

\textit{Mean-Field Theory.}
Motivated by the well-documented success of the simplest BCS-BEC 
crossover approach developed for one-band Fermi gases across a 
magnetic Feshbach resonance~\cite{giorgini08}, here we introduce 
a BCS-like intra-band order parameter
$
\Delta_i = - \sum_j V_{ij} \langle b_{j \mathbf{0}} \rangle
$
for each band, where $\langle \cdots \rangle$ is a thermal average within
the mean-field pairing approximation. Therefore, the complex parameters 
$\Delta_1$ and $\Delta_2$ are uniform in space since we set 
$\mathbf{q} = \mathbf{0}$, and they in general need to be determined 
together with the corresponding number equations
$
N_i = \sum_{\sigma \mathbf{k}} \langle c_{i \sigma \mathbf{k}}^\dagger  c_{i \sigma \mathbf{k}} \rangle.
$
After some straightforward algebra, the grand potential becomes~\cite{josephson}
$
\Omega_{mf} =
\sum_{i \mathbf{k}}  [\xi_{i \mathbf{k}} - E_{i \mathbf{k}}
- 2 T \ln (1 + e^{-E_{i \mathbf{k}}/T} )]
+ (V_{22}|\Delta_1|^2 + V_{11}|\Delta_2|^2
- V_{12} \Delta^*_1\Delta_2 - V_{21} \Delta^*_2\Delta_1)/(V_{11}V_{22}-V_{12}V_{21}),
$
and the resultant self-consistency equations can be compactly 
written as
\begin{align}
\label{eqn:op}
\Delta_i & = \sum_{j \mathbf{k}} V_{ij} \frac{\Delta_j}{2E_{j \mathbf{k}}} \tanh \left(\frac{E_{j \mathbf{k}}}{2T} \right), \\
N_i & = \sum_{\mathbf{k}} \left[ 1 - \frac{\xi_{i \mathbf{k}}}{E_{i \mathbf{k}}} \tanh \left(\frac{E_{i \mathbf{k}}}{2T} \right)\label{eqn:ne}
\right],
\end{align}
where
$
E_{i \mathbf{k}} = \sqrt{\xi_{i \mathbf{k}}^2 + |\Delta_i|^2}
$
is the energy of the quasi-particle excitations in the $i$th band, $T$ is the 
temperature, and the Boltzmann constant $k_B$ is set to unity. 
The set of coupled equations given above is the generalization of the 
usual one-band crossover approach to the case of multi-band systems,
and we hope that this description remain sufficient in understanding 
(at least qualitatively) some of the low-$T$ properties of alkaline-earth 
Fermi gases across an orbital Feshbach resonance in the entire 
parameter regime of interest. 

In addition to the trivial ($\Delta_{1} = \Delta_{2} = 0$) one, Eq.~(\ref{eqn:op}) 
allows for two nontrivial solutions with 
$
\Delta_i = |\Delta_i| e^{i\varphi_i},
$
such that the grand potential is minimized or maximized either by 
the in-phase ($\varphi_1 = \varphi_2$) or the out-of-phase ($\varphi_1 = \varphi_2 + \pi$) 
one depending on whether the Josephson-type coupling $V_{12} = V_{21}$ 
is positive or negative as long as the stability condition 
$V_{11}V_{22} > V_{12}V_{21}$ manifests. As discussed below, the 
so-called $\pi$-phase solution is directly associated with the recently-realized 
orbital Feshbach resonance in a $^{173}$Yb Fermi gas~\cite{pagano15, hofer15}. 
Thus, the sign of the inter-band coupling determines the relative 
phases of the stable (ground state) and unstable (excited state) solutions for a 
generic two-band model. In contrast to the relative phase, the overall phase 
$\varphi_1 + \varphi_2$ is not a physical observable, and its random value is 
chosen spontaneously at a given realization. Note that the $V_{12} = V_{21} = 0$ 
limit is not entirely trivial because while $\varphi_1$ and $\varphi_2$ are 
completely uncoupled, \textit{i.e.}, they are not physical observables, 
$|\Delta_1|$ and $|\Delta_2|$ are still coupled when $\delta \ne 0$ 
even in the $V_{11} = V_{22}$ case considered in this paper.
In addition, while $\mu$ is varied in such a way to fix the overall number 
of particles $N = N_1+N_2$ at a particular value (see the section on 
numerical results), the relative band population is determined self-consistently 
for a given $\delta$.

\textit{Analytical Limits.}
It is desirable to gain intuitive understanding of the analytically-tractable limits 
as much as possible before going through the cumbersome details of the 
fully-numerical calculations. For instance, we reach the following conclusions 
by simply analyzing the general structure of Eq.~(\ref{eqn:op}).

\begin{itemize}

\item $g_\pm \to 0$ or equivalently $1/a_{s\pm} \to -\infty$ limit. This leads to a
unique in-phase/out-of-phase solution with $|\Delta_1| = |\Delta_2|$ determined by
$
1 = g_\mp \sum_{i \mathbf{k}} \tanh \left[E_{i \mathbf{k}}/(2T) \right]/(4E_{i \mathbf{k}}).
$

\item $g_+ \to g_- = g$ or equivalently $a_{s+} = a_{s-}$ limit. Since $V_{12} = V_{21} = 0$ 
in this limit, $\varphi_1$ and $\varphi_2$ are not physical observables, and
the non-trivial values for $|\Delta_1|$ and $|\Delta_2|$ are determined by
$
1 = g \sum_{\mathbf{k}} \tanh \left[E_{i \mathbf{k}}/(2T) \right]/(2E_{i \mathbf{k}}).
$
This suggests that either $|\Delta_2|$ (ground state) or $|\Delta_1|$ (excited state)
vanishes unless $\delta \approx 0$, in such a way that the particles gradually 
transfer from the upper (lower) to the lower (upper) band with increasing 
$g$ in the ground (excited) state.

\item $\delta \to 0$ limit. Since $V_{11} = V_{22}$ in this paper, the degenerate 
bands become completely symmetric in this limit, giving rise to $|\Delta_1| = |\Delta_2|$ 
and $E_{1 \mathbf{k}} = E_{2 \mathbf{k}} = E_{\mathbf{k}}$. This leads to a 
unique in-phase/out-of-phase solution for $g_\mp > g_\pm$ (or equivalently 
$1/a_{s\mp} > 1/a_{s\pm}$), where
$
1 = g_\mp \sum_{\mathbf{k}} \tanh \left[E_{\mathbf{k}}/(2T) \right]/(2E_{\mathbf{k}}).
$

\item $\delta \to \infty$ limit. If $|\Delta_{1(2)}| \gg |\Delta_{2(1)}|$ then we obtain
$\Delta_{2(1)}/\Delta_{1(2)} = (g_\mp - g_\pm)/(g_- + g_+)$. This leads to a 
unique in-phase (out-of-phase) / out-of-phase (in-phase) solution for 
$g_\mp > g_\pm$. It is clear that these limits are relevant in the neighbourhood
of $g_+ \approx g_-$ (or equivalently $a_{s+} \approx a_{s-}$).

\item $\Delta_{1 (2)} \ne 0$ but $\Delta_{2 (1)} = 0$ limit. This requires $g_+ = g_- = g$ 
with
$
1 = g \sum_{\mathbf{k}} \tanh \left[E_{1 (2) \mathbf{k}}/(2T) \right]/(2E_{1 (2) \mathbf{k}}),
$
in agreement with the previous limits discussed above.

\end{itemize}

These analyses clearly show that the allowed nontrivial solutions are unique only 
in some special limits, and we rely heavily on these analytical limits while 
characterizing the numerically-obtained results in general, as thoroughly 
discussed next.

\textit{Numerical Results.}
In order to obtain cutoff-independent results for $|\Delta_1|$, $|\Delta_2|$ and $\mu$,
we solve the self-consistency Eqs.~(\ref{eqn:op}) and~(\ref{eqn:ne}) with a large 
$\mathbf{k}$-space cutoff $k_0 = 100 k_F$, where the Fermi momentum $k_F$ 
determines the overall density of particles via the usual free-particle relation 
$n = N/\Omega = k_F^3/(3\pi^2)$, written for the lowest band. The corresponding 
Fermi energy $\varepsilon_F = k_F^2/(2m)$ is assumed to be less than 
$\delta/2$ as illustrated in Fig.~\ref{fig:bands}. Since we iterate $\mu$ to keep 
$N$ fixed at this value in all of our numerical calculations, here we present 
only the relative population $|N_1-N_2|/N$.

\begin{figure}[htb]
\centerline{\scalebox{0.41}{\includegraphics{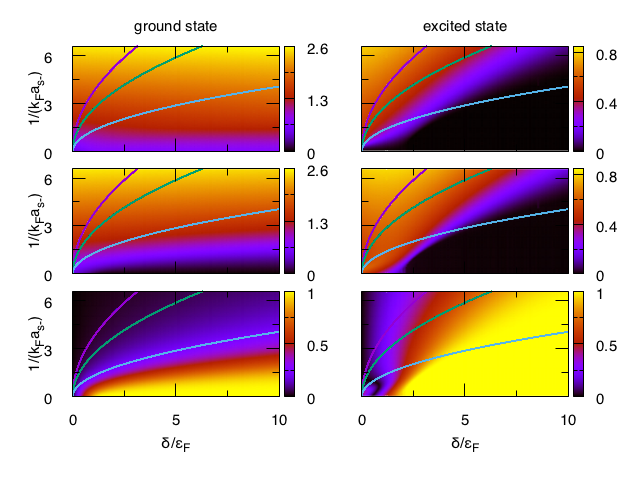}}}
\caption{\label{fig:Yb} (Color online) \textit{$^{173}$Yb Fermi gas}. 
The colored maps of the magnitudes of the intra-band order parameters 
$|\Delta_1|/\epsilon_F$ (top) and $|\Delta_2|/\epsilon_F$ (middle) as well as 
the relative band population $|N_1-N_2|/N$ (bottom) are shown for 
$a_{s+} \approx 1900a_0$ and $a_{s-} \approx 200a_0$. 
The left and right panels correspond, respectively, to the in-phase and 
out-of-phase solutions discussed in the text. In addition, 
$\delta = \delta_{res}$ (purple), $\delta = 2\delta_{res}$ (green) and 
$\delta = \delta_0$ (light blue) contours are explicitly shown.
}
\end{figure}

In Fig.~\ref{fig:Yb}, we show $|\Delta_1|$, $|\Delta_2|$ and $|N_1-N_2|/N$ for 
the scattering parameters $a_{s\pm}$ of a $^{173}$Yb Fermi 
gas~\cite{pagano15, hofer15}. Depending on the local density of a given 
experimental setup, one can easily extract the relevant local 
parameters from this figure. For instance, typical atomic systems 
have densities of order $n \approx 10^{14} \textrm{cm}^{-3}$, 
corresponding to the dimensionless parameters $1/(k_F a_{s+}) \approx 0.69$, 
$1/(k_F a_{s-}) \approx 6.58$ and $\delta_{res}/\varepsilon_F \approx 3.14$.
At resonance, we find $\Delta_1 \approx 2.55 \epsilon_F$, 
$\Delta_2 \approx 2.51 \epsilon_F$, $\mu \approx -46 \epsilon_F$ and $N_1-N_2 = 0.027N$ 
for the in-phase (ground-state), and $\Delta_1 \approx 0.65 \epsilon_F$, 
$\Delta_2 \approx -0.76 \epsilon_F$, $\mu \approx 0.34 \epsilon_F$ and 
$N_1-N_2 = 0.41N$ for the out-of-phase (excited-state) solution. 
Away from the resonance, Fig.~\ref{fig:Yb} clearly shows that while the latter 
solution depends strongly on $\delta$, the former solution does not.
For instance, when $\delta = 2\delta_{res}$, we find $\Delta_1 \approx 2.57 \epsilon_F$, 
$\Delta_2 \approx 2.48 \epsilon_F$, $\mu \approx -46 \epsilon_F$ and 
$N_1-N_2 = 0.053N$ for the in-phase, and $\Delta_1 \approx 0.38 \epsilon_F$, 
$\Delta_2 \approx -0.53 \epsilon_F$, $\mu \approx 0.79 \epsilon_F$ and 
$N_1-N_2 = 0.79N$ for the out-of-phase solution. This suggests that the 
excited-state solution is driven mainly by the orbital Feshbach resonance.

To further support this inference, we note that if $a_{s\pm} > a_{s\mp} > 0$ 
then the anti-symmetric and symmetric superpositions of the order parameters
$
\Delta_{\pm} = (\Delta_2 \mp \Delta_1)/2
$
are associated, respectively, with the shallow ($a_{s\pm}$) and deep 
($a_{s\mp}$) two-body bound states. In addition, since the shallow one is solely 
responsible for the orbital Feshbach resonance, these bound states 
correspond, respectively, to the excited- and ground-state solutions 
discussed above. In contrast, if only one of $a_{s\pm}$ is positive then 
the corresponding bound state $a_{s\pm}$, and hence the combination 
$\Delta_\pm$, is associated with the orbital Feshbach resonance as 
long as $a_{s2} > 0$. Thus, as a rule of thumb, we conclude that the 
formation of bound states across an orbital Feshbach resonance is 
associated with the $\max |\Delta_\pm|$ superposition whenever 
$a_{s\pm} > |a_{s\mp}|$, \textit{i.e.}, the out-of-phase/in-phase solution. 
This analysis again suggests that it is the excited-state (out-of-phase) 
solution that is predominantly characterized by the orbital Feshbach 
resonance in a $^{173}$Yb Fermi gas.

\begin{figure}[htb]
\centerline{\scalebox{0.41}{\includegraphics{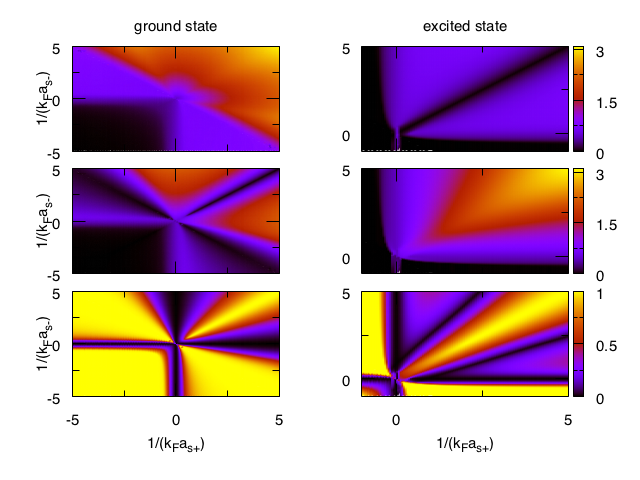}}}
\caption{\label{fig:generic} (Color online) \textit{Resonant Fermi gas}.
The colored maps of the magnitudes $|\Delta_1|/\epsilon_F$ (top), 
$|\Delta_2|/\epsilon_F$ (middle) and $|N_1-N_2|/N$ (bottom) are 
shown for $\delta = 1/(m a_{s2}^2)$, so that the gas is at resonance 
in the regions with $a_{s+} + a_{s-} > 0$. The left (right) panel 
corresponds to the in-phase/out-of-phase (out-of-phase/in-phase) 
solution depending on $1/a_{s-} \gtrless 1/a_{s+}$,
}
\end{figure}

In Fig.~\ref{fig:generic}, we show $|\Delta_1|$, $|\Delta_2|$ and $|N_1-N_2|/N$ 
for a generic Fermi gas with arbitrary $a_{s \pm}$ values when 
$\delta = 1/(m a_{s2}^2)$. Since a wide-range of parameter region 
satisfying the condition $a_{s2} > 0$ corresponds to a resonant Fermi gas
in general, this figure may serve as a first-hand guide for future experiments 
with other species of alkaline-earth atoms. In particular, we find that the 
excited-(ground-)state solution strengthens (weakens) as $a_{s+} \to a_{s-} > 0$ 
in agreement with our analytical analyses given above, suggesting that the 
alkaline-earth Fermi gases with $a_{s+} \approx a_{s-} > 0$ are best 
candidates for investigating many-body phenomena across an orbital 
Feshbach resonance~\cite{trap}.

\textit{Conclusions.}
To conclude, there is arguably no doubt that the exact analogy between 
the Hamiltonian of alkaline-earth Fermi gases across an orbital Feshbach 
resonance~\cite{zhang15, xu16} and of two-band $s$-wave superconductors 
with contact interactions~\cite{suhl59} opens yet a new frontier for studying 
exotic condensed-matter phenomena in atomic systems. In particular, 
the key role played by the inter-band pair scattering between two intra-paired 
bands is played in atomic systems by the intra-orbital scattering lengths 
between the open and closed channels, and the physics is similar in many 
ways to intrinsic Josephson effect between two condensates~\cite{josephson, current}. 
This makes atomic settings an ideal test bed for studying the competition between 
the in-phase and out-of-phase solutions of the superfluid order parameters 
and their fluctuations. For instance, in addition to the familiar phonon-like 
in-phase fluctuations of the relative phase, \textit{i.e.}, the massless 
Goldstone mode, it may be possible to study the long-sought exciton-like 
out-of-phase fluctuations of the relative phase, \textit{i.e.}, the massive 
Leggett mode~\cite{leggett66, iskin05, bittner15, liu16}.

\textit{Acknowledgments.} 
This work is supported by the funding from T\"{U}B$\dot{\mathrm{I}}$TAK Grant 
No. 1001-114F232 and the BAGEP award of the Turkish Science Academy, 
and the author thanks Hui Zhai for his correspondence on past and forthcoming 
developments.


\begin{thebibliography}{99}
\bibitem{footHam} This representation makes it explicit that the many-body 
description of an alkaline-earth $^{173}$Yb Fermi gas across an orbital 
Feshbach resonance~\cite{zhang15, pagano15, hofer15, xu16} is exactly 
analogous to that of simplest two-band $s$-wave superconductors with contact 
interactions~\cite{suhl59}.
\bibitem{josephson} Note that the inter-band coupling gives rise to a 
Josephson-type contribution to the grand potential,
$-2V_{12}|\Delta_1 \Delta_2| \cos(\varphi_1-\varphi_2) /(V_{11}V_{22}-V_{12}^2)$,
which depends explicitly on the relative phase of the intra-band order parameters.
\bibitem{current} \textit{The AC Josephson effect}. Accordingly, it can be shown that 
the relative population evolve in time $t$ as
$
\Delta N = \Delta N_0 - 16V_{12}|\Delta_{10} \Delta_{20}| \cos(\Delta \varphi_0) 
\sin^2(\delta t/4)/[\delta (V_{11}V_{22}-V_{12}^2)],
$
where $\Delta N_0 = N_{20}-N_{10}$, $\Delta \varphi_0 = \varphi_{20}-\varphi_{10}$, 
etc. are the equilibrium ($t = 0$) values. Note that this expression is strictly valid for 
parameter regimes when $\Delta N \approx \Delta N_0$ and 
$|\Delta_i| \approx |\Delta_{i0}|$. Thus, for $k_0/k_F \to \infty$, the population 
imbalance $P = (N_2-N_1)/N$ evolves, respectively, as
$
P = P_0 \mp  3\pi |\Delta_{10} \Delta_{20}|/(\delta \epsilon_F) 
[1/(k_F a_{s-}) - 1/(k_F a_{s+})] \sin^2(\delta t/4)
$
for the in-phase and out-of-phase solutions. Using the parameters given in the 
text, this reduces to
$
P = P_0 \mp  55.5|\Delta_{10} \Delta_{20}|/(\delta \epsilon_F) \sin^2(\delta t/4)
$
for an $^{173}$Yb Fermi gas.
\bibitem{trap} Due to the presence of a confining trap potential, the Fermi gas 
forms a shell structure where the superfluid core is surrounded by a normal 
gas that is induced by $\delta \ne 0$ even at $T = 0$ (not shown).
\end{thebibliography}
\end{document}